\documentclass[twocolumn,amsmath,amssymb]{revtex4}

\def\gsim{ \lower .75ex \hbox{$\sim$} \llap{\raise .27ex \hbox{$>$}} }

\def\lsim{ \lower .75ex \hbox{$\sim$} \llap{\raise .27ex \hbox{$<$}} }

\def\IZ{\relax\ifmmode\mathchoice
{\hbox{\cmss Z\kern-.4em Z}}{\hbox{\cmss Z\kern-.4em Z}}
{\lower.9pt\hbox{\cmsss Z\kern-.4em Z}} {\lower1.2pt\hbox{\cmsss
Z\kern-.4em Z}}\else{\cmss Z\kern-.4em Z}\fi}
\def\IR{\relax{\rm I\kern-.18em R}}

\def\one{{\hbox{ 1\kern-.8mm l}}}

\newlength{\bredde}
\def\slash#1{\settowidth{\bredde}{$#1$}\ifmmode\,\raisebox{.15ex}{/}
\hspace*{-\bredde} #1\else$\,\raisebox{.15ex}{/}\hspace*{-\bredde}
#1$\fi}

\newcommand  {\Rbar} {{\mbox{\rm$\mbox{I}\!\mbox{R}$}}}

\newsavebox{\zzzbar}
\sbox{\zzzbar}
  {\setlength{\unitlength}{0.9em}
  \begin{picture}(0.6,0.7)
  \thinlines
  \put(0,0){\line(1,0){0.6}}
  \put(0,0.75){\line(1,0){0.575}}
  \multiput(0,0)(0.0125,0.025){30}{\rule{0.3pt}{0.3pt}}
  \multiput(0.2,0)(0.0125,0.025){30}{\rule{0.3pt}{0.3pt}}
  \put(0,0.75){\line(0,-1){0.15}}
  \put(0.015,0.75){\line(0,-1){0.1}}
  \put(0.03,0.75){\line(0,-1){0.075}}
  \put(0.045,0.75){\line(0,-1){0.05}}
  \put(0.05,0.75){\line(0,-1){0.025}}
  \put(0.6,0){\line(0,1){0.15}}
  \put(0.585,0){\line(0,1){0.1}}
  \put(0.57,0){\line(0,1){0.075}}
  \put(0.555,0){\line(0,1){0.05}}
  \put(0.55,0){\line(0,1){0.025}}
  \end{picture}}

\newcommand{\ena}{\end{eqnarray}}
\newcommand{\beqa}{\begin{eqnarray}}
\newcommand{\eeqa}{\end{eqnarray}}
\newcommand{\bea}{\begin{eqnarray}}
\newcommand{\eea}{\end{eqnarray}}

\newcommand{\be}{\begin{equation}}
\newcommand{\ee}{\end{equation}}

\usepackage{graphicx}

\def\be{\begin{equation}}
\def\ee{\end{equation}}
\def\beq{\begin{eqnarray}}
\def\eeq{\end{eqnarray}}

\def\({\left (}
\def\){\right )}
\def\[{\left [}
\def\[{\right ]}

\def\ba{\begin{eqnarray}}
\def\ea{\end{eqnarray}}

\input amssym.def
\input amssym.tex

\usepackage{graphicx}

\usepackage{dcolumn}

\usepackage{bm}

\begin{document}

\title{From Big Crunch to Big Bang with AdS/CFT}

\author{Neil Turok,$^1$ Ben Craps,$^{2,3}$ Thomas Hertog$^{3,4}$}

\affiliation{
{\em $^1$ DAMTP, CMS, Wilberforce Road, Cambridge, CB3 0WA, UK and\\
African Institute for Mathematical Sciences, 6-8 Melrose Rd, Muizenberg 7945, RSA}
\vskip .5mm
{\em$^2$ Theoretische Natuurkunde, Vrije Universiteit Brussel,
Pleinlaan 2, B-1050 Brussels, Belgium}
\vskip .5mm 
{\em$^3$International Solvay Institutes, Boulevard du Triomphe,
ULB--C.P.231, B-1050 Brussels, Belgium}
\vskip .5mm 
{\em$^4$APC, Universit\'e Paris 7, 10 rue A.Domon et L.Duquet, 75205 Paris, France}
} 
\begin{abstract} 
The AdS/CFT correspondence is used to describe five-dimensional cosmology with a big crunch singularity in terms of super-Yang-Mills theory on $\Rbar\times S^3$ deformed by a potential which is unbounded below. Classically, a Higgs field in the dual theory rolls to infinity in finite time.  But since the $S^3$ is finite, the unstable mode spreads quantum mechanically and the singularity is resolved when self-adjoint boundary conditions are imposed at infinity. Asymptotic freedom of the coupling governing the instability gives us computational control and the quantum spreading provides a UV cutoff on particle creation. The bulk interpretation of our result is a quantum transition from a big crunch to a big bang. An intriguing consequence of the near scale-invariance of the dual theory is that a nearly scale-invariant spectrum of stress-energy perturbations is automatically generated in the boundary theory. We comment on implications for more realistic cosmologies.
\end{abstract}


\maketitle

A fundamental challenge facing string and M-theory is that of dealing with the big bang singularity. Not only did the visible cosmos emanate from it, so too did the extra dimensions whose configuration determines the laws of particle physics and key parameters such as the dark energy density. Hence it is vital both to particle physics and to cosmology that we resolve the initial singularity. 

It is natural to turn for guidance to the AdS/CFT correspondence,~\cite{Maldacena:1997re} which has emerged as an extremely powerful tool for studying quantum gravity. According to AdS/CFT, a gravitational theory in an asymptotically Anti-de Sitter (AdS) spacetime is dual to a quantum field theory on its conformal boundary. In this holographic setup, the radial dimension of {\it space} is emergent, but {\it time}  is not: there is a global time which plays an entirely conventional role in the boundary theory. 

In this Letter, we use AdS/CFT to study a big crunch cosmology, where the bulk singularity corresponds to the roll of a boundary scalar field to infinity in finite time. Several ingredients allow us to describe the dynamics in detail. First, the relevant coupling in the boundary theory is asymptotically free, rendering perturbation theory reliable. Second, the boundary evolution is smooth and ultralocal, even nonlinearly, as the singularity is approached. Finally, the boundary has finite volume, so the homogeneous mode of the unstable field undergoes quantum spreading. With self-adjoint boundary conditions imposed at large field values, the wavefunction reflects off infinity with vanishing weight there, providing an automatic UV cutoff on particle creation. Remarkably, the boundary theory develops nearly scale-invariant fluctuations as a consequence of its (slightly broken) conformal invariance, suggesting a new, natural mechanism for generating cosmological perturbations. A detailed account of this work is given in Ref.~\cite{CHT}.

Ten-dimensional type IIB supergravity compactified on $S^5$ can be consistently truncated, in several inequivalent ways, to five dimensional gravity coupled to a single scalar field $\varphi$, describing a quadrupole distortion of the $S^5$.\cite{Gunaydin:1985cu} For example, truncating to an $SO(5)$-invariant scalar $\varphi$ yields the Lagrangian 
\be
\frac{1}{2}R-\frac{1}{2} (\nabla\varphi)^2+\frac{1}{4} R_{AdS}^{-2}\left(15e^{2\gamma \varphi}+10 e^{-4\gamma \varphi}-e^{-10\gamma\varphi} \right)
\ee
with $\gamma=(2/15)^{1/2}$. The static AdS$_5$ vacuum occurs at $\varphi=0$, where the scalar field mass saturates the Breitenlohner-Freedman (BF) bound $m^2=-4R_{AdS}^{-2}$. We are interested in time-dependent supergravity solutions which evolve into a big crunch singularity in the future (and possibly out of a big bang singularity in the past). In all solutions asymptotic to AdS$_5$, with $ds^2=R_{AdS}^2 [-(1+r^2)dt^2+dr^2/(1+r^2)+r^2d\Omega_3^2]$, $\varphi$ tends to $\alpha r^{-2}\ln r +\beta r^{-2}$, with $\alpha$ and $\beta$ functions of the boundary coordinates, at large $r$. Specifying the dynamics requires a boundary condition on timelike infinity relating $\alpha$ to $\beta$. The supersymmetric, AdS-invariant choice is $\alpha=0$: but $\alpha=f \beta$, with $f>0$, allows smooth, asymptotically AdS data to evolve to a big crunch singularity in finite global time. 

For $f=0$, the dual field theory is $SU(N)$ gauge theory with ${\cal N}=4$ supersymmetry on $\Rbar \times S^3$. The bulk scalars saturating the BF bound couple to quadratic traces of Higgs scalars traceless on their $SO(6)$ indices. For example, the $SO(5)$-invariant bulk scalar $\varphi$, couples to ${\cal O}\sim N^{-1}{\rm Tr}[\Phi_1^2-(1/5) \sum_{i=2}^6\Phi_i^2]$, where the trace is on $SU(N)$ indices. The function $\alpha$ appearing in the boundary condition plays the role of a source for ${\cal O}$ in the boundary theory, while $\beta$ corresponds to the expectation value of ${\cal O}$. Boundary conditions with $f>0$ correspond to a deformation of the CFT: $ S\rightarrow S+\int f {\cal O}^2$,\cite{Witten:2001ua} {\it i.e.}, a scalar potential driving ${\cal O}$ to infinity in finite time. Similar cosmological duals were discussed in Refs.~\cite{Hertog:2004rz} but the results were inconclusive because the dual 3d theories are poorly understood.\cite{Elitzur:2005kz} In the example chosen here, the dual 4d theory is a renormalizable theory with a perturbative regime. Furthermore,
because of the ``wrong'' sign of the deformation, the coupling $f$ is asymptotically free. It follows that quantum corrections are under excellent perturbative control for large ${\cal O}$, {\it i.e.}, near the singularity. The large $N$ effective potential reads~\cite{Schwimmer:2003eq}
\be
V=-{\cal O}^2/\ln({\cal O}/\tilde{M}^2) \equiv - f_{\cal O} {\cal O}^2.
\label{epot}
\ee
Through dimensional transmutation, $f$ has been replaced by an arbitrary mass scale $\tilde{M}$.\cite{Coleman:1973jx} If $f_{\cal O}$ is small at one value of ${\cal O}$, it is even smaller for larger values, so one can be sure that quantum corrections do not turn the potential around. At small 't~Hooft coupling, we have control over the gauge theory. According to the AdS/CFT relation $g_{t} \equiv g_{YM}^2N=(R_{AdS}/\ell_s)^4$, the bulk is then in a stringy regime. The supergravity regime corresponds to large 't~Hooft coupling $g_t$ in the gauge theory. However, Witten's discussion~\cite{Witten:2001ua} indicates (\ref{epot}) is still valid. Hence the unbounded negative potential is a robust feature at small or large 't~Hooft coupling. 

Since $V$ is unbounded below, it is not obvious the field theory makes sense. A key point is that the $S^3$ is finite, so the unstable homogeneous mode is quantum mechanical. Any initial wavepacket spreads instantaneously to infinity, so one simply cannot discuss the quantum evolution without imposing a boundary condition at large ${\cal O}$, chosen to ensure the Hamiltonian is self-adjoint.  In  quantum mechanics this construction is known as the ``self-adjoint extension.''\cite{Reed:1975uy,Carreau90} We study its field theory implementation and argue furthermore that perturbation theory is reliable for computing the amplitude to obtain almost all final values of the homogeneous mode.

The self-adjoint extension is usually formulated~\cite{Carreau90} as a boundary condition on energy eigenstates. We reformulate it using the time-dependent semi-classical expansion, which becomes exact near the singularity.\cite{CHT} The Schr\"odinger equation is solved as an expansion in $\hbar$, $\Psi = (A(x,t)+\hbar \dots) e^{i S(x,t)/\hbar +\dots} $ where $S(x,t)$ is the action for a complex classical solution ending at real $x$ at time $t$. The initial condition for a Gaussian wavepacket centered on $x_c$, with momentum $p_c$, is $x+2 i (\Delta x)^2 p/\hbar= x_c+2 i (\Delta x)^2 p_c/\hbar$, where $\Delta x$ is the initial spread in $x$. The imaginary part of the classical solution is thereby related to the quantum spreading of the wavefunction. 

For a potential like $V(x)=-(1/4) \lambda x^{4}$, invariant under reflecting $x$ through infinity in the complex $x$-plane, a wavefunction satisfying a self-adjoint boundary condition may be constructed via the method of images. The ``image'' part of the wavefunction is obtained from a complex solution with $(x_c,p_c)\rightarrow -(x_c,p_c)$ but the same final value of $x$. The complete semiclassical wavefunction is
\be
\Psi^\alpha(x,t)= A_1(x,t) e^{i S_1/\hbar} + e^{-2i \alpha} A_2(x,t) e^{i S_2/\hbar},
\label{psieqa}
\ee
where $S_1$ is the action for the first complex solution, $S_2$ is that for the ``image'' solution, $A_i$ are the corresponding amplitude factors and $0<\alpha<\pi$ is an arbitrary phase specifying the particular self-adjoint extension. At large $x$, we find the $x$-dependence to be
\be
\Psi^\alpha(x,t)\rightarrow  A(t)\, x^{-1} \cos\left(\sqrt{\lambda} x^{3}/(3\sqrt{2})+\alpha\right),
\label{asymform}
\ee
that of the self-adjoint energy eigenstates~\cite{Carreau90}: thus $\Psi^\alpha$ lies in the domain of the self-adjoint extension. The physical interpretation is that a right-moving wavepacket disappearing at infinity is always accompanied by a left-moving wavepacket appearing at infinity, with phase $2 \alpha$. To describe a bounce, we start with a Gaussian wavepacket rolling downhill, dominated by the first term in (\ref{psieqa}). The wavefunction reflects from infinity and rolls back uphill in a similar wavepacket, dominated by the second, ``image'' term. The phase $\alpha$ then nearly cancels in the final probability density. 

We can extend this technique to the dual field theory. We focus on the steepest negative direction of $V$, parameterized as $\Phi_1 = \phi\, U$, with $U$ a normalized $SU(N)$ generator, ${\rm Tr}(U^2)=1$, and $\Phi_i=0$, $i=2\dots 6$.  The effective potential becomes
\be 
V \sim R_{AdS}^{-2} \phi^2 - \lambda_\phi \phi^4, \, \lambda_\phi= 1/\left(N^2 {\rm ln}(\phi/(N M))\right) \equiv N^{-2} l^{-1},
\label{epot1}
\ee
ignoring numerical factors, with $M\equiv \tilde{M} N^{-1/ 2}$.
We assume $M R_{AdS} \ll 1$, so $l$ is large for all $\phi$ beyond the barrier. In this case, many quantities of interest can be expanded in inverse powers of $l$. Consider the classical solution starting with zero kinetic and potential energy, at $\phi \sim \lambda_\phi^{-1/2} R^{-1}_{AdS}$.  As $\phi$ rolls to large values, it approaches the scaling solution $\phi \sim \lambda_\phi^{-1/2} |t|^{-1}$ plus $l^{-1}$ corrections. Defining $\chi= \int_\phi^\infty d \phi \, \lambda_\phi^{-1/2} \phi^{-2}  \sim \lambda_\phi^{-1/2} \phi^{-1}(1+{1\over 2}l^{-1}+\dots)$, the solution is just $\chi=|t|$. Initially, $\chi \sim R_{AdS}$: the time to bounce off infinity and return is $\sim 2 R_{AdS}$. Now consider a Gaussian wavepacket in $\phi$, centered on the above  solution. Given an initial spread $\Delta \phi_i$, we can use $\Delta \pi_{\phi,i} \sim R_{AdS}^3 \Delta \dot{\phi}_i \sim \hbar/\Delta \phi_i$ to estimate the final spread, $\Delta \chi_f \sim ( \Delta \chi_i^2 + R_{AdS}^2 (\Delta \dot{\chi}_i)^2)^{1/2}$. This is minimized by $\Delta \chi_{i,min} \sim \lambda_\phi^{1/2} R_{AdS}\ll \chi_c$, or $\Delta \phi_{i,min} \sim R_{AdS}^{-1}\ll \phi_c$, where $\phi_c$ is the center of the initial wavepacket and $\chi_c$ the corresponding value of $\chi$. We shall write the spread of the initial wavepackets we consider as $W\Delta \phi_{i,min}$.

To describe the bouncing wavepacket we must find the appropriate complex classical solution. For times $t\approx +R_{AdS}$, the wavefunction is dominated by the image solution. Its initial condition is $\phi + 2 i \hbar^{-1} \pi_\phi (\Delta \phi)^2 =-\phi_c$, re-expressed in $\chi$. With the final condition $\chi=r_f$, real, corresponding to $\phi=\phi_f$, the solution is fully determined. To a first approximation $l$ may be treated as constant and energy conservation leads to $\dot{\chi}=(1+e\chi^4)^{1/2}$ with $e$ a complex constant.  Setting $\chi=r+is$, one finds $r$ is close to the real, zero-energy scaling solution $r= t$, and $s\approx (s_i/2) \left( 1- (r/r_f)^5 \right).$
Up to numerical factors, $s_i \sim W^2(r_f-r_{Cl})$, where $r_{Cl}$ is the classical value of $r$ at the final time, and thus positive (negative) if the argument of the wavefunction $r_f$ is ahead of (behind) the classical bouncing solution. 

Including the running of the coupling $\lambda_\phi$ complicates the analysis because the logarithm introduces a branch cut at $\phi=\infty$. However, imposing a UV cutoff $\Lambda$, the effective potential $V$ is meromorphic for $|\phi|>\Lambda$ and hence the method of images still works. We carefully compute the relevant classical solution at large, finite $\Lambda$, following it onto the appropriate Riemann sheet before finally taking $\Lambda$ to infinity. The final wavefunction is $\Lambda$-independent and defines a completely unitary evolution.\cite{CHT}

In describing the bouncing homogeneous mode, we have so far ignored particle creation. We can check this is consistent by including the inhomogeneous modes in our two complex classical solutions. Each mode is placed in the incoming adiabatic vacuum state by imposing the initial condition that it is purely positive frequency. We then solve for the inhomogeneous modes to quadratic order, determining the amplitude of the outgoing negative frequency component. Finally, we compare the energy density in quantum-created particles with the kinetic or potential energy in the homogeneous mode.  As we shall see, in a particular parameter regime, backreaction is negligible. 

When the modes of interest have wavelengths $\ll R_{AdS}$, the curvature of the $S^3$ can be safely neglected. The equation of motion for fluctuations in $\phi$ is 
\be 
\ddot{\delta \phi} = \left(-k^2 +6 \chi^{-2} (1+{5\over 12}l^{-1}+\dots )\right)\delta \phi 
\equiv - \omega^2(t) \delta \phi.
\label{dpeq} 
\ee
The semiclassical expansion gives the wavefunction for the mode exactly to quadratic order,
\be
\Psi(\delta \phi,t) \sim e^{i \frac{{\cal S}_{Cl}}{ \hbar}}, \qquad {\cal S}_{Cl} = \frac{1}{2} (\dot{R}_{in}/ R_{in}) (\delta \phi)^2,
\label{evalhamact} 
\ee
with $R_{in}\rightarrow e^{i\omega t}, kt \ll -1$, being the incoming positive frequency mode. The number of created particles at the final time $t_f$ is
\be
\langle n \rangle =
 \frac{(\dot{R}_{in}-i\omega(t_f) R_{in})(\dot{R}_{in}^*+i\omega(t_f) R_{in}^*) }{ (-2 i \omega (t_f)) (R_{in}^*\dot{R}_{in} - R_{in} \dot{R}_{in}^*)},
\label{expecn}
\ee  
where $w(t_f)$ is real. 
Now, a very important fact comes into play. Near the singularity, the complex solution for the homogeneous mode is well-approximated by $\chi=t + i \epsilon$, where $\epsilon = s_i/2$ is a constant of order  $W^2 \lambda_\phi^{1/2} R_{AdS}$, for typical final values of the homogeneous background field $\phi_f$. For most values, therefore, the dangerous $\chi^{-2}$ term in (\ref{dpeq}) is actually bounded and smooth at $t=0$. Hence quantum particle production is exponentially suppressed for $k >|\epsilon|^{-1} \sim W^{-2}  \lambda_\phi^{-1/2} R^{-1}_{AdS} \sim W^{-2} N l^{1/2} R_{AdS}^{-1}$.  

At leading (zeroth) order in $l^{-1}$, the solution $R_{in}$ of (\ref{dpeq}) is $(t+i\epsilon)^{1/2} H_{5/2}(k(t+i\epsilon))$ with the property that a positive frequency incoming mode evolves to a positive frequency outgoing mode. This behavior was first noticed in field theory on compactified Milne spacetime, where $\epsilon$ was used to define a choice of analytic continuation.~\cite{Tolley:2002cv} Here, $\epsilon$ is a physical parameter determined by the final value of the homogeneous ``background,'' {\it i.e.}, the argument of the wavefunction, and except when $\epsilon$ vanishes, the evolution of linearized fluctuations is unambiguous. Solving to next order in $l^{-1}$, we find the positive frequency incoming mode evolves to $\sim e^{ikt} -i\pi e^{-ikt} e^{-2k|\epsilon|}/\ln(k/M)$ in the outgoing adiabatic regime, $kt\gg 1$, $t \sim R_{AdS} \ll M^{-1}$, giving the energy density in quantum-created $\delta \phi$ particles 
\be
\rho_{c,\delta \phi}= \int \frac{d^3{\bf k}}{(2 \pi)^3} k \frac{\pi^2}{\ln^2(k/M)} e^{-4k |\epsilon|}\sim \frac{N^4}{W^8R_{AdS}^4},
\label{parts}
\ee
where the logarithm enters as the value of $l$ when a mode leaves the adiabatic regime, when $|kt| \sim 1$.  

We now consider quantum production of other modes. Assuming $\Phi_1$ points in general position in $SU(N)$ space, {\it i.e.}, $SU(N)$ is broken to $U(1)^{N-1}$, $N$ gauge, Higgs and Fermi particles remain light. The other Higgs and gauge particles acquire masses ${\cal M} \sim g_{YM} \phi$ with $\dot{\cal M}/{\cal M}^2 \sim \lambda_\phi^{1/2} /g_{YM} \sim (g_t l N)^{-1/2} \ll 1$ at large $l$ and $N$, so the corresponding positive frequency mode evolution is adiabatic and particle production is exponentially suppressed.  

The light Higgs particles acquire mass from the term $\lambda_\phi {\rm Tr}(\Phi_1^2) {\rm Tr}(\Phi_i^2)$, $i=2,\dots,6$. Their mode functions obey  $\ddot \Phi_i = -k^2 \Phi_i-(2/ 5) (1+ l^{-1}+O(l^{-2})) \Phi_i/\chi^2$. To zeroth order in $l^{-1}$, $R_{in}$ is again a Hankel function and yields no particle production. To first order in $l^{-1}$, up to a numerical factor we find an energy density $N$ times greater than (\ref{parts}). 

The massless gauge bosons have no classical coupling to $\phi$. However, a coupling arises at one loop. The effective Lagrangian for the massless gauge fields is $-(1/4) F^2 \left(1 + g_{YM}^2\sum_{i=S,F,V} c_i {\rm Tr}\ln ({\cal M}_i^2/\mu^2)\right)$, with $\mu$ a renormalization scale and the trace taken over the heavy scalars, fermions and gauge bosons, with mass matrices ${\cal M}_i$.\cite{ball} In the undeformed theory, the correction vanishes by supersymmetry. The deformation alters ${\cal M}_S^2 \sim g^2 \Phi_1^2 \rightarrow g^2 \Phi_1^2 + \lambda_\phi \Phi_1^2$, producing a finite, $\mu$-independent correction $\sim N^{-1} l^{-1} F^2 $. The equation of motion for the massless gauge boson modes yields a mode mixing coefficient $\sim N^{-1} e^{-2k |\epsilon|}/\ln(k/M)^2$ across the bounce. Compared to (\ref{parts}), the produced energy density is suppressed by one power of $N$ and two powers of $\ln(k/M)$. 

A conservative criterion for backreaction to be small is that the homogeneous mode should keep enough energy to roll back up the hill to nearly its starting value,
$\phi_{start}\sim R_{AdS}^{-1} N \left(\ln(\phi_{start}/NM)\right)^{1/2}\approx R_{AdS}^{-1} N |\ln(MR_{AdS})|^{1/2}$.
The minimal value of $\phi$ reached after the bounce is obtained by setting $V(\phi_{min})=-\rho_c$. Demanding that $\phi_{min}-\phi_{start}\ll\phi_{start}$ in order to have small backreaction, one finds the condition $\rho_c\ll N^2|\ln(MR_{AdS})|/R_{AdS}^4$; as the largest contribution to $\rho_c$ comes from light Higgs particles with $\rho_c\sim N^5/W^8R_{AdS}^4$, this becomes
\be\label{logN3}
|\ln(MR_{AdS})|\gg N^3/W^8.
\ee
So, for example, if we choose $W\sim N^{3/8}$, backreaction is under control 
as long as $|\ln(MR_{AdS})|\gg 1$, for final values of $\phi$ carrying most of the probability. This value of $W$ corresponds to a spread in the time delay $\epsilon\sim N^{-1/4} |\ln(MR_{AdS})|^{-1/2} R_{AdS}$, parametrically small compared to the duration $\sim 2 R_{AdS}$ of the bounce. Hence our wavepacket is well-described by a classical bouncing solution, with small backreaction. The fraction of the probability associated with final $\phi$ values for which the backreaction is large (and hence our calculation breaks down) is 
$\sim |\ln(MR_{AdS})|^{-1/4} $. Hence we conclude that when the logarithm is large, the most probable outcome is a bounce. 


While the main emphasis of our work is to establish the possibility of a nonsingular bounce, 
the fluctuations in the rolling field $\phi$ have a further remarkable consequence: they give rise to a nearly scale-invariant spectrum of stress-energy perturbations in the dual theory. As $\phi$ rolls, its fluctuation $\delta \phi$ acquires an increasingly tachyonic mass, causing modes of successively higher $k$ to cease oscillating and enter a ``growing mode" solution, just as in the ekpyrotic mechanism.\cite{Khoury:2001wf} In that case, the scale-invariant spectrum generated is the result of the classical scale-invariance of a scalar field with a negative exponential potential under rescaling spacetime coordinates and shifting the field $\phi \rightarrow \phi + {\rm const}$. Here, the field theory Lagrangian is also unstable and classically scale-invariant, with the scaling symmetry rescaling $\phi$ according to its canonical dimension. The classical instability leads naturally to a nearly scale-invariant spectrum of stress-energy fluctuations on the boundary. Quantum effects break the scale symmetry but with small anomalous dimensions because $\lambda_\phi$ is asymptotically free. A straightforward but lengthy computation of the two-point correlator for the ``improved'' stress-energy tensor yields for the dimensionless fluctuation $\delta \equiv \delta \rho/(P+\rho) $, with $\rho=T^{00}$ and $P=T^{ii}/3$ as usual, the two-point correlator $\langle \delta (r,t)\delta(0,t)\rangle \sim  N^{-2} (\ln Mr)^{-2} (\ln Mt)^{-1} F(r/t)$ with $F$ a dimensionless function. The only violations of naive (engineering dimensions) scale-invariance are due to the logarithms and they are small at weak coupling. The stress-energy perturbations are (i) naturally small, being suppressed by powers of $1/N$ and $\lambda_\phi$, (ii) approximately scale-invariant, with a slightly red tilt due to asymptotic freedom, (iii) scalar, (iv) adiabatic and (v) nearly Gaussian in character, to leading order in $\lambda_\phi$ and $1/N$. This is a tantalizing result. To be sure, we have only so far computed boundary correlators and hence not the quantities of direct interest in the bulk cosmology. However, since the only scale which enters the translation is $R_{AdS}$ and we are interested in modes with $k \gg R_{AdS}^{-1}$, and since scale-invariance is ``holographic'' in the sense that the fluctuations on a planar subspace of a space with scale-invariant fluctuations are scale-invariant, it is plausible that the bulk cosmological perturbations will turn out to be scale-invariant.  (For related calculations in $O(1,3)$-invariant backgrounds see Ref.~\cite{Gratton:1999ya}.)

A more realistic cosmology requires a setup with a four dimensional bulk, like $AdS_4 \times S^7$.\cite{Hertog:2004rz} The conformal boundary is $\Rbar \times S^2$ and the dual theory is believed to be the IR limit of a $2+1$-dimensional, large $N$ gauge theory. As in our case, the dual theory is expected to be classically conformal-invariant and unstable: the  model dynamics proposed in Ref.~\cite{Hertog:2004rz} consists of a scalar field $\phi$ with a deformation potential $- f \phi^6$. In this model, $f$ is  again asymptotically free. While precise calculations are currently beyond reach, it is not unreasonable to expect qualitatively similar behavior to that found here. If so, and if the model can be embedded in a bigger picture allowing for a positive dark energy phase, this mechanism may provide a rather compelling explanation for the origin of the observed density variations in the universe.

We thank N.~Arkani-Hamed, N.~Dorey, D.~Gross, G.~Horowitz, J.~Maldacena, E.~Martinec, R.~Myers, S.~Sethi, D.~Tong and H.~Verlinde for encouragement and discussions. N.T. acknowledges support from STFC (UK) and the Centre for Theoretical Cosmology (CTC) in Cambridge. The work of B.C. was supported in part by the Belgian Federal Science Policy Office through the Interuniversity Attraction Pole IAP VI/11, by the European Commission FP6 RTN programme MRTN-CT-2004-005104 and by FWO-Vlaanderen through project G.0428.06. T.H. thanks the KITP in Santa Barbara for support.  B.C. and T.H. thank the CTC for hospitality.

\nopagebreak

\end{document}